\documentclass[conference]{IEEEtran}

\usepackage[T1]{fontenc}
\newtheorem{lemma}{Lemma}
\newtheorem{theorem}{Theorem}
\usepackage{graphicx}
\usepackage{cite}
\usepackage{float}
\usepackage{epstopdf} 
\usepackage{amsmath}
\usepackage{amssymb}

\begin{document}

\title{Uplink Spectral Efficiency of Massive MIMO with Spatially Correlated Rician Fading}

\author{\IEEEauthorblockN{\"Ozgecan \"Ozdogan, Emil Bj\"ornson, Erik G. Larsson}
\IEEEauthorblockA{Department of Electrical Engineering (ISY), Link\"oping University, Sweden\\
Email: \{ozgecan.ozdogan, emil.bjornson, erik.g.larsson\}@liu.se}}

\maketitle

\begin{abstract}

This paper considers the uplink (UL) of a multi-cell Massive MIMO (multiple-input multiple-output) system with spatially correlated Rician fading channels. The channel model is composed of a deterministic line-of-sight (LoS) path and a stochastic non-line-of-sight (NLoS) component describing a spatially correlated multipath environment. We derive the statistical properties of the minimum mean squared error (MMSE) and least-square (LS) channel estimates for this model. Using these estimates for maximum ratio (MR) combining, rigorous closed-form UL spectral efficiency (SE) expressions are derived. Numerical results show that the SE is higher when using the MMSE estimator than the LS estimator, and the performance gap increases with the number of antennas. Moreover, Rician fading provides higher achievable SEs than Rayleigh fading since the LoS path improves the sum SE. 
\end{abstract}


\IEEEpeerreviewmaketitle

\section{Introduction}

Massive MIMO  is the key technology for increasing the SE in future cellular networks, by virtue of beamforming and spatial multiplexing \cite{Larsson2014}. A Massive MIMO BS is equipped with a massive number (e.g., a hundred)  of steerable antennas and is able of serving tens of user equipments (UEs) simultaneously. The canonical form of Massive MIMO operates in time-division duplex (TDD) mode and acquires the channel state information (CSI), necessary for UL receive combining and downlink (DL) precoding, from UL pilot signaling \cite{EmilsBook}. 

The achievable SEs of Massive MIMO systems with imperfect CSI have been rigorously characterized and optimized for fading channels modeled by either spatially uncorrelated~\cite{Marzetta2016a} or spatially correlated~\cite{EmilsBook} Rayleigh fading. Communication with fading-free LoS propagation has also be treated \cite{EmilsBook,Yang2017}. However, practical channels can consist of both a deterministic LoS path and small-scale fading caused by multipath propagation, which can be jointly modeled by the Rician fading model \cite{Tse2005a}.

The performance of Massive MIMO with Rician fading channels is much less analyzed than with Rayleigh fading. The single-cell case was studied in \cite{Zhang2014,Kong2015,Hu2017} under the assumption of spatially uncorrelated Rician fading channels and zero-forcing (ZF) processing. Approximate SE expressions for the UL and DL were provided in \cite{Zhang2014} and \cite{Kong2015,Hu2017}, respectively. The multi-cell case was studied in \cite{Zhao2016,Wu2017,Sanguinetti2017}, assuming spatially uncorrelated Rician fading within each cell and spatially uncorrelated Rayleigh fading across cells. Approximate SE expressions were derived in the UL with ZF combining \cite{Wu2017} and in the DL with ZF \cite{Zhao2016} or regularized ZF precoding  \cite{Sanguinetti2017}. Note that these are the papers that consider imperfect CSI, which is the practically relevant case, while prior works assuming perfect CSI can be found in the reference lists of \cite{Zhang2014,Kong2015,Hu2017,Zhao2016,Wu2017,Sanguinetti2017}.

There are three major limitations of the prior works. First, the fading was modeled as spatially uncorrelated, although practical channels are correlated, due to the finite number of scattering clusters \cite{EmilsBook}. Second, the inter-cell channels were modeled by Rayleigh fading, although it happens that a UE has LoS paths to multiple BSs (e.g., in parks, dense small-cell deployments, or when serving unmanned aerial vehicles (UAVs)). Third, only approximate SE expressions were derived in closed form, which only provide insights into special operational regimes, such as having asymptotically many antennas. In this paper, we address these shortcomings:

\begin{itemize}
\item We consider a multi-cell scenario with spatially correlated Rician fading channels between every pair of BSs and UEs. Previously, this channel model has only been used for single-cell scenarios with perfect CSI \cite{Zhang2013,Tataria2017}.

\item We derive the MMSE and LS channel estimates and characterize their statistics. Using these estimates for MR combining, we compute rigorous closed-form UL SEs.

\item We compare the SEs with MMSE and LS estimation numerically, considering both Rician and Rayleigh fading.

\end{itemize}

\section{Channel and System Model}

We consider a Massive MIMO system with $L$ cells, where the $j$th BS has $M_j$ antennas and serves $K_j$ single-antenna UEs. The channel responses remain constant over a coherence block of $\tau_c$ samples and the channel realizations are independent between any pair of blocks. The value of $\tau_c$ is determined by the carrier frequency and external factors such as the propagation environment and UE mobility \cite[Ch.~2]{Marzetta2016a}. In this paper, we focus on the UL, using  $\tau_p$ samples for UL pilot signals and $\tau_u=\tau_c-\tau_p$  samples for UL data transmission.

The channel between UE $k$ in cell $l$ and the BS in cell $j$ is denoted by $\mathbf{h}^j_{lk} \in \mathbb{C}^{M_j}$. The superscript of $\mathbf{h}^j_{lk}$ indicates the BS index and the subscript identifies the cell and index of the UE. We consider spatially correlated Rician fading channels where $\mathbf{h}^j_{lk}$, $\forall j,l \in 1,\dots,L$ and $\forall k \in 1,\dots,K$, is a realization of the circularly symmetric complex Gaussian distribution
\begin{equation}\label{eq1}
	\mathbf{h}^j_{lk} \sim \mathcal{N}_\mathbb{C}\left(  \bar{\mathbf{h}}^j_{lk}, \mathbf{R}^j_{lk} \right)
\end{equation}
where the mean $ \bar{\mathbf{h}}^j_{lk} \in \mathbb{C}^{M_j}$ corresponds to the LoS component and $\mathbf{R}^j_{lk} \in \mathbb{C}^{M_j \times M_j}$ is the positive semi-definite covariance matrix describing the spatial correlation of the NLoS components. The small-scale fading is described by the Gaussian distribution whereas  $\mathbf{R}^j_{lk}$ and $\bar{\mathbf{h}}^j_{lk}$ model the macroscopic propagation effects, including the antenna gains and radiation patterns at the transmitter and receiver. The analysis in this paper holds for any values of these parameters, while a specific model is considered in Section~\ref{sec:numerical_results}.

\section{Channel Estimation} \label{sec4}
Each BS requires channel state information (CSI) for receive processing. Therefore, $\tau_p$ samples are reserved for performing UL pilot-based channel estimation in each coherence block, giving room for $\tau_p$ mutually orthogonal pilot sequences. 
These pilot sequences are allocated to different UEs and the same sequences are reused by UEs in multiple cells. The deterministic pilot sequence of UE $k$ in cell $j$ is denoted by $\boldsymbol{\phi}_{jk} \in \mathbb{C}^{\tau_p}$ and  $\|\boldsymbol{\phi}_{jk} \|^2= \tau_p$.
We define the set
\begin{equation}
	\mathcal{P}_{jk} = \left\lbrace (l,i): \boldsymbol{\phi}_{li} = \boldsymbol{\phi}_{jk}, l=1,\dots,L, i=1,\dots,K_l \right\rbrace 
\end{equation}
with indices of all UEs that utilize the same pilot sequence as UE $k$ in cell $j$ (including the UE itself). The received pilot signal $\mathbf{Y}^p_{j} \in \mathbb{C}^{M_j \times \tau_p }$   at BS $j$ is
\begin{equation}
	\mathbf{Y}^p_{j} = \sum_{k=1}^{K_j} \sqrt{p_{jk}} \mathbf{h}^j_{jk} \boldsymbol{\phi}^T_{jk} + \mathop{\sum_{l=1 }}^{L}_{l \neq j} \sum_{i=1}^{K_l} \sqrt{p_{li}} \mathbf{h}^j_{li} \boldsymbol{\phi}^T_{li} + \mathbf{N}^p_j
\end{equation}
where $\mathbf{N}^p_j \in \mathbb{C}^{M_j\times \tau_p}$ has independent and identically distributed circularly symmetric complex Gaussian entries with zero mean and variance $\sigma^2_{\mathrm{ul}}$. 
To estimate the channel $\mathbf{h}^j_{li}$, BS~$j$ multiplies $\mathbf{Y}^p_{j}$ with the UE's pilot sequence $\boldsymbol{\phi}^*_{li}$ to obtain 
\begin{equation}\label{sec3eq1}
	\mathbf{y}^p_{jli}= \mathbf{Y}^p_{j}  \boldsymbol{\phi}^*_{li}=\sqrt{p_{li}} \tau_p \mathbf{h}^j_{li} + \!\sum_{(l',i') \in \mathcal{P}_{li} \backslash (l,i)} \!\!\!\! \sqrt{p_{l'i'}} \tau_p \mathbf{h}^j_{l'i'} +\mathbf{N}^p_j \boldsymbol{\phi}^*_{li}.
\end{equation}
The processed received pilot signal $\mathbf{y}^p_{jli} \in \mathbb{C}^{M_j}$ is a sufficient statistics for estimating $\mathbf{h}^j_{li}$.

\subsection{MMSE Channel Estimator}
Based on the processed received pilot signal $\eqref{sec3eq1}$, the BS can apply MMSE estimation to obtain an estimate of ${\mathbf{h}}^j_{li}$ as shown in the following lemma. Notice that Bayesian estimators require that the statistical distributions (the mean vector and covariance matrices) are known. These can be estimated using the sample mean and sample covariance matrices in practice.
\begin{lemma}
The MMSE estimate of the channel from BS $j$  to UE $i$ in cell $l$ is
	\begin{equation}\label{mmse1}
		\hat{\mathbf{h}}^j_{li} = \bar{\mathbf{h}}^j_{li} + \sqrt{p_{li}}  \mathbf{R}^j_{li} \boldsymbol{\Psi}^j_{li} \left( \mathbf{y}^p_{jli} - \bar{\mathbf{y}}^p_{jli}  \right) 
	\end{equation}
	where $\bar{\mathbf{y}}^p_{jli}=\sum_{(l',i') \in \mathcal{P}_{li} } \sqrt{p_{l'i'}} \tau_p \bar{\mathbf{h}}^j_{l'i'} $	and 
	\begin{align}
		\boldsymbol{\Psi}^j_{li}={\tau_p} \mathrm{Cov}\left\lbrace {\mathbf{y}^p_{jli}}\right\rbrace^{-1} \!=\!\Bigg(  \sum_{(l',i') \in \mathcal{P}_{li} } \!\!{p_{l'i'}} \tau_p \mathbf{R}^j_{l'i'} +  \sigma^2 \mathbf{I}_{M_j} \Bigg)^{\!-1}\!\!.
	\end{align}
	The estimation error $\tilde{\mathbf{h}}^j_{li} = \mathbf{h}^j_{li} - \hat{\mathbf{h}}^j_{li}$ has the covariance matrix
	\begin{equation}
		\mathbf{C}^j_{li}=  \mathbf{R}^j_{li} - p_{li} \tau_p \mathbf{R}^j_{li} \boldsymbol{\Psi}^j_{li} \mathbf{R}^j_{li}
	\end{equation}
	and the mean-squared error is $\mathrm{MSE}= \mathbb{E}\lbrace \|  \mathbf{h}^j_{li} - \hat{\mathbf{h}}^j_{li}\|^2 \rbrace  =\mathrm{tr}( \mathbf{C}^j_{li} ) $. The MMSE estimate  $\hat{\mathbf{h}}^j_{li}$ and the estimation error $\tilde{\mathbf{h}}^j_{li} $ are independent random variables and distributed as
	\begin{equation}  \label{eq:statistics-MMSE-estimate}
		\hat{\mathbf{h}}^j_{li} \sim \mathcal{N}_\mathbb{C}\left( \bar{\mathbf{h}}^j_{li},  \mathbf{R}^j_{li} - \mathbf{C}^j_{li} \right),
	\end{equation}
	\begin{equation}
		\tilde{\mathbf{h}}^j_{li} \sim \mathcal{N}_\mathbb{C}\left( \mathbf{0}_M, \mathbf{C}^j_{li} \right).
	\end{equation}
\end{lemma}
\begin{IEEEproof}
The proof follows from the standard MMSE estimation of Gaussian random variables \cite{KayBookESt,EmilsBook}.
\end{IEEEproof}

Note that the estimation error covariance matrix $\mathbf{C}^j_{li}$ does not depend on the mean values. In other words, the estimation error is not affected by LoS components, under the assumption that the mean values are known. Moreover, the channel estimates of UEs in the set $\mathcal{P}_{li}$ are not independent, since they use the same pilot sequence. This is known as pilot contamination. UE $(j,k) \in \mathcal{P}_{li}$ has the channel estimate
\begin{equation}
	\hat{\mathbf{h}}^j_{jk} = \bar{\mathbf{h}}^j_{jk} + \sqrt{p_{jk}}  \mathbf{R}^j_{jk} \boldsymbol{\Psi}^j_{li} \left( \mathbf{y}^p_{jli} - \bar{\mathbf{y}}^p_{jli}  \right) 
\end{equation}
and it is correlated with $\hat{\mathbf{h}}^j_{li}$ in \eqref{mmse1} since $\mathbf{y}^p_{jli} $ appears in both expressions and $\boldsymbol{\Psi}^j_{li}=\boldsymbol{\Psi}^j_{jk}$. We will utilize the distributions of the channel estimates and estimation errors in Section~\ref{section:SE} when analyzing the UL SE.

\subsection{LS Channel Estimator}
If the BS has no prior information regarding $ \mathbf{R}^j_{li}$ and $\bar{\mathbf{h}}^j_{li}$, the LS estimator can be utilized to get an estimate of the propagation channel $\mathbf{h}^j_{li}$. The LS estimate is defined as the value of $\hat{\mathbf{h}}^j_{li}$ that minimizes $\| \mathbf{y}^p_{jli} - \sqrt{p_{li}} \tau_p\hat{\mathbf{h}}^j_{li} \|^2 $, that is 
\begin{equation}\label{ls1}
\hat{\mathbf{h}}^j_{li} = \frac{1}{\sqrt{p_{li}} \tau_p} \mathbf{y}^p_{jli}.
\end{equation}
\begin{lemma}
	The LS estimator and estimation error are correlated random variables and distributed as
	\begin{align} 
	\hat{\mathbf{h}}^j_{li} &\sim \mathcal{N}_\mathbb{C}\left( \frac{1}{\sqrt{p_{li}} \tau_p} \bar{\mathbf{y}}^p_{jli},  \ \ \frac{1}{p_{li} \tau_p} (\boldsymbol{\Psi}^j_{li})^{-1} \right) \\
	\tilde{\mathbf{h}}^j_{li} &\sim \mathcal{N}_\mathbb{C}\left( \bar{\mathbf{h}}^j_{li}\! -\! \frac{1}{\sqrt{p_{li}} \tau_p} \bar{\mathbf{y}}^p_{jli} ,  \frac{1}{p_{li} \tau_p} (\boldsymbol{\Psi}^j_{li})^{-1} \!-\mathbf{R}^j_{li}   \!\right).
	\end{align}
\end{lemma}	
\begin{IEEEproof}
The proof is omitted due to space limitations, but follows from computing first- and second-order moments.
\end{IEEEproof}

Note that the estimation errors have non-zero mean in this case, since the statistics are not utilized by the LS estimator.

\begin{figure*}[!h]
	\normalsize
	\setcounter{equation}{18}
	
	\begin{align}\label{sec4eq4}
	&\mathbb{E}\left\lbrace \left| {\mathbf{v}}_{jk}^H \mathbf{h}^j_{li}  \right| ^2 \right\rbrace = p_{jk} \tau_p \mathrm{tr}\left( \mathbf{R}^j_{li} \mathbf{R}^j_{jk} \boldsymbol{\Psi}^j_{jk} \mathbf{R}^j_{jk}\right) + p_{jk} \tau_p ( \bar{\mathbf{h}}^j_{li})^H    \mathbf{R}^j_{jk} \boldsymbol{\Psi}^j_{jk} \mathbf{R}^j_{jk} \bar{\mathbf{h}}^j_{li} +  ( \bar{\mathbf{h}}^j_{jk})^H    \mathbf{R}^j_{li} \bar{\mathbf{h}}^j_{jk} +  \left|  (\bar{\mathbf{h}}^j_{jk})^H    \bar{\mathbf{h}}^j_{li}   \right| ^2 \nonumber \\  
	&+\begin{cases}
	p_{jk} p_{li}  \tau^2_p \left|  \mathrm{tr}\left( \mathbf{R}^j_{li} \boldsymbol{\Psi}^j_{jk} \mathbf{R}^j_{jk}\right) \right| ^2 +  2 \sqrt{ p_{jk} p_{li}}\tau_p \mathrm{Re}\left\lbrace\mathrm{tr}\left(    \mathbf{R}^j_{li} \boldsymbol{\Psi}^j_{jk} \mathbf{R}^j_{jk}  \right)  ( \bar{\mathbf{h}}^j_{li})^H \bar{\mathbf{h}}^j_{jk}  \right\rbrace   &  (l,i) \in \mathcal{P}_{jk}   \\
	0 &  (l,i) \notin \mathcal{P}_{jk}
	\end{cases}
	\end{align}
	\hrulefill
	\vspace*{0pt}
\end{figure*}
\begin{figure*}[h]
	\normalsize
		\setcounter{equation}{20}
	\begin{equation}\label{xi1}
	{\xi}^{\mathrm{ul}}_{li}= \frac{p_{jk} \tau_p \mathrm{tr}\left( \mathbf{R}^j_{li} \mathbf{R}^j_{jk} \boldsymbol{\Psi}^j_{jk} \mathbf{R}^j_{jk}\right) + p_{jk} \tau_p \left( \bar{\mathbf{h}}^j_{li}\right)^H    \mathbf{R}^j_{jk} \boldsymbol{\Psi}^j_{jk} \mathbf{R}^j_{jk} \bar{\mathbf{h}}^j_{li} +  \left( \bar{\mathbf{h}}^j_{jk}\right)^H    \mathbf{R}^j_{li} \bar{\mathbf{h}}^j_{jk} + \left|  (\bar{\mathbf{h}}^j_{jk})^H    \bar{\mathbf{h}}^j_{li}   \right| ^2}{ p_{jk} \tau_p \mathrm{tr}\left( \mathbf{R}^j_{jk} \boldsymbol{\Psi}^j_{jk} \mathbf{R}^j_{jk}\right) +\| \bar{\mathbf{h}}^j_{jk}   \|^2} 
	\end{equation}
	\hrulefill
	\vspace*{0pt}
\end{figure*}
\begin{figure*}[h]
	\normalsize
		\setcounter{equation}{21}
	\begin{equation}\label{coherentInterference}
	{\Gamma}^{\mathrm{ul}}_{li}=    \frac{p_{jk} p_{li}  \tau^2_p\left|  \mathrm{tr}\left( \mathbf{R}^j_{li} \boldsymbol{\Psi}^j_{jk} \mathbf{R}^j_{jk}\right) \right| ^2 +  2 \sqrt{ p_{jk} p_{li}}\tau_p \mathrm{Re}\left\lbrace \mathrm{tr}\left(    \mathbf{R}^j_{li} \boldsymbol{\Psi}^j_{jk} \mathbf{R}^j_{jk}  \right) \left( \bar{\mathbf{h}}^j_{li}\right)^H \bar{\mathbf{h}}^j_{jk}  \right\rbrace }{p_{jk} \tau_p \mathrm{tr}\left( \mathbf{R}^j_{jk} \boldsymbol{\Psi}^j_{jk} \mathbf{R}^j_{jk}\right) +\| \bar{\mathbf{h}}^j_{jk}   \|^2} 
	\end{equation}
	\hrulefill
	\vspace*{0pt}
\end{figure*}

\section{Spectral Efficiency with MR Combining}
\label{section:SE}
\setcounter{equation}{13}
During UL data transmission, the received signal $\mathbf{y}_j \in \mathbb{C}^{M_j}$ at BS $j$ is
\begin{align}\label{s1eq1}
	\mathbf{y}_j = \displaystyle\sum_{k=1}^{K_j} \mathbf{h}^j_{jk} s_{jk} + \mathop{\sum_{l=1 }}^{L}_{l \neq j}\sum_{i=1}^{K_l} \mathbf{h}^j_{li} s_{li} + \mathbf{n}_j
\end{align}
where $\mathbf{n}_j \sim \mathcal{N}_\mathbb{C}\left(  \mathbf{0}_{M_j}, \sigma^2_{\mathrm{ul}} \mathbf{I}_{M_j}    \right) $ is the additive noise. The UL signal from UE $k$ in cell $l$ is denoted by $s_{lk} \in \mathbb{C}$ and has power $p_{lk} = \mathbb{E}\left\lbrace |s_{lk}|^2\right\rbrace $. The first term in \eqref{s1eq1} is the desired signal and the latter terms denote interference and noise, respectively. 

BS~$j$ selects the receive combining vector $\mathbf{v}_{jk} \in \mathbb{C}^{M_j}$ based on its CSI and multiplies it with $\mathbf{y}_j$ to separate the desired signal of its UE $k$ from interference. 
As in \cite[Th.~4.4]{EmilsBook}, the ergodic UL capacity of UE $k$ in cell $j$ is lower bounded by 
\begin{equation} \label{eq:SEexpression}
	\mathrm{SE}^{\mathrm{ul}}_{jk}=\frac{\tau_u}{\tau_c}\log_2\left(1 + \gamma^{\mathrm{ul}}_{jk} \right) \  \ \mathrm{[bit/ s/ Hz]}
\end{equation}
where the effective SINR $\gamma^{\mathrm{ul}}_{jk}  $ is equal to
\begin{equation}\label{sec4eq1}
	\frac{p_{jk} |  \mathbb{E}\lbrace \mathbf{v}^H_{jk} \mathbf{h}^j_{jk}  \rbrace   | ^2}{\displaystyle\sum_{l=1}^{L}\sum_{i=1}^{K_l}   p_{li}  \mathbb{E}\lbrace | \mathbf{v}^H_{jk} \mathbf{h}^j_{li}  | ^2 \rbrace  - p_{jk} |  \mathbb{E}\lbrace \mathbf{v}^H_{jk} \mathbf{h}^j_{jk}  \rbrace   | ^2  +\sigma^2_{\mathrm{ul}}  \mathbb{E}\lbrace  \| \mathbf{v}_{jk} \|^2 \rbrace } 
\end{equation}
where the expectations are with respect to all sources of randomness. Since $\mathrm{SE}^{\mathrm{ul}}_{jk}$ is below the capacity, it is an achievable SE.
The  SINR $\gamma^{\mathrm{ul}}_{jk}$ can be computed numerically for any combining scheme and channel estimator. We will show that it can be computed in closed form when using MR combining.

\subsection{Spectral Efficiency with the MMSE estimator}
If the MMSE estimator in \eqref{mmse1} is used, we obtain a closed-form expression for the SE in \eqref{eq:SEexpression} as in the next theorem.

 \begin{theorem}
 If MR  combining with $\mathbf{v}_{jk} = \hat{\mathbf{h}}^j_{jk}$ is used based on the MMSE estimator, then 
	\begin{equation}\label{sec4eq2}
		\mathbb{E}\left\lbrace \mathbf{v}^H_{jk} \mathbf{h}^j_{jk}  \right\rbrace = p_{jk} \tau_p \mathrm{tr}\left( \mathbf{R}^j_{jk} \boldsymbol{\Psi}^j_{jk} \mathbf{R}^j_{jk}\right) +\| \bar{\mathbf{h}}^j_{jk}   \|^2
	\end{equation}
	\begin{equation}\label{sec4eq3}
		\mathbb{E}\left\lbrace  \| \mathbf{v}_{jk} \|^2 \right\rbrace  = p_{jk} \tau_p \mathrm{tr}\left( \mathbf{R}^j_{jk} \boldsymbol{\Psi}^j_{jk} \mathbf{R}^j_{jk}\right) +\| \bar{\mathbf{h}}^j_{jk}   \|^2
	\end{equation}
and $\mathbb{E}\lbrace | {\mathbf{v}}_{jk}^H \mathbf{h}^j_{li}  | ^2 \rbrace$ is given in \eqref{sec4eq4}, at the top of this page. Plugging these expressions into the SINR in $\eqref{sec4eq1}$ yields 
\setcounter{equation}{19}
\begin{equation} \label{MMSEeq1}
	\gamma^{\mathrm{ul}}_{jk} =\frac{ p_{jk}^2 \tau_p \mathrm{tr}\left( \mathbf{R}^j_{jk} \boldsymbol{\Psi}^j_{jk} \mathbf{R}^j_{jk}\right) + p_{jk}\| \bar{\mathbf{h}}^j_{jk}   \|^2}{\displaystyle\sum_{l=1}^{L}\sum_{i=1}^{K_l}   p_{li} {\xi}^{\mathrm{ul}}_{li}  +  \displaystyle\sum_{(l,i) \in \mathcal{P}_{jk} \backslash (j,k)} p_{li} {\Gamma}^{\mathrm{ul}}_{li}  - p_{jk}\nu_{jk}^{\mathrm{ul}} + \sigma^2_{\mathrm{ul}} }
\end{equation}
where  ${\nu}_{jk}^{\mathrm{ul}}=\frac{\| \bar{\mathbf{h}}^j_{jk}   \|^4}{p_{jk} \tau_p \mathrm{tr}\left( \mathbf{R}^j_{jk} \boldsymbol{\Psi}^j_{jk} \mathbf{R}^j_{jk}\right) +\| \bar{\mathbf{h}}^j_{jk}   \|^2}$, ${\xi}^{\mathrm{ul}}_{li}$ and ${\Gamma}^{\mathrm{ul}}_{li}$ are defined in \eqref{xi1} and \eqref{coherentInterference}, at the top of this page, and these terms correspond to LoS-related interference, non-coherent interference, and coherent interference, respectively.

\end{theorem}
\begin{IEEEproof}
	The proof is omitted due to space limitations, but follows from direct, tedious computation of expectations. 
	\end{IEEEproof}
	
	The rigorous closed-from SINR expression in \eqref{MMSEeq1}  provides important and exact insights into the behaviors of Rician fading Massive MIMO systems. The signal terms in the numerator depend on the estimation quality and the LoS component. The former is reduced by pilot contamination, since $ p_{jk}^2 \tau_p \mathrm{tr}\left( \mathbf{R}^j_{jk} \boldsymbol{\Psi}^j_{jk} \mathbf{R}^j_{jk}\right) = p_{jk} \mathrm{tr}\left( \mathbf{R}^j_{jk} - \mathbf{C}^j_{jk}\right) $, which contains the covariance matrix of the estimate in \eqref{eq:statistics-MMSE-estimate}.
	
	In the denominator, the relation between the covariance matrices $\mathbf{R}^j_{jk}$ and $\mathbf{R}^j_{li}$, and  the inner product of LoS components $\bar{\mathbf{h}}^j_{jk} $ and $\bar{\mathbf{h}}^j_{li} $ determine how large the interference terms are. If the covariance matrices span different subspaces, or one has small eigenvalues, there will be little interference from the NLoS propagation. Similarly, there is little interference from the LoS propagation when $\bar{\mathbf{h}}^j_{jk} $ and $\bar{\mathbf{h}}^j_{li}$ are nearly orthogonal.
	
The non-coherent interference term ${\xi}^{\mathrm{ul}}_{li}$ does not increase with $M_j$, unless $\bar{\mathbf{h}}^j_{jk} $ and $ \bar{\mathbf{h}}^j_{li} $ are nearly parallel vectors. The coherent interference term ${\Gamma}^{\mathrm{ul}}_{li}$ involves the pilot-contaminating UEs which are $(l,i) \in \mathcal{P}_{jk} \backslash (j,k)$ and it grows linearly with $M_j$. The term ${\nu}_{jk}^{\mathrm{ul}}$ grows with $M_j$ and  depends on the norm of desired UE's LoS component. 
\begin{figure*}[t]
	\normalsize
	\setcounter{equation}{24}
	\begin{align}\label{lseq3}
	&{{p_{jk}} \tau^2_p}\mathbb{E}\left\lbrace |\mathbf{v}^H_{jk} \mathbf{h}^j_{li} |^2 \right\rbrace ={{p_{jk}} \tau^2_p} \chi_{li}^{\mathrm{ul,ls}}  =   \tau_p\mathrm{tr}\left( \mathbf{R}^j_{li}(\boldsymbol{\Psi}^j_{jk})^{-1} \right) +2\sqrt{p_{li}} \tau_p  \mathrm{Re}\left\lbrace  (\bar{\mathbf{y}}^p_{jjk})^H \bar{\mathbf{h}}^j_{li} \mathrm{tr}\left(  \mathbf{R}^j_{li} \right) +  (\bar{\mathbf{y}}^p_{jjk})^H \mathbf{R}^j_{li} \bar{\mathbf{h}}^j_{li} \right\rbrace    \\
	&\!+\!\begin{cases}
	(\bar{\mathbf{y}}^p_{jjk})^H \mathbf{R}^j_{li}\bar{\mathbf{y}}^p_{jjk} + \tau_p (\bar{\mathbf{h}}^j_{li})^H (\boldsymbol{\Psi}^j_{jk})^{-1}\bar{\mathbf{h}}^j_{li} + |(\bar{\mathbf{y}}^p_{jjk})^H \bar{\mathbf{h}}^j_{li}|^2    &  (l,i) \!\notin \mathcal{P}_{jk} \nonumber \\ 
	p_{li} \tau^2_p  |  \mathrm{tr}(  \mathbf{R}^j_{li})| ^2 +  \bar{\mathbf{x}}^H_{jk}\mathbf{R}^j_{li}\bar{\mathbf{x}}_{jk}+ \tau_p (\bar{\mathbf{h}}^j_{li})^H (\boldsymbol{\Omega}^j_{jk})^{-1} \bar{\mathbf{h}}^j_{li}  + |\bar{\mathbf{x}}^H_{jk} \bar{\mathbf{h}}^j_{li}|^2  + p_{li} \tau^2_p  \|\bar{\mathbf{h}}^j_{li} \|^4 + 2\sqrt{p_{li}} \tau_p  \mathrm{Re}\left\lbrace   \bar{\mathbf{x}}^H_{jk}   \bar{\mathbf{h}}^j_{li} \| \bar{\mathbf{h}}^j_{li} \|^2 \right\rbrace \!\!  &  (l,i) \!\in \mathcal{P}_{jk}
	\end{cases}
	\end{align}
	
	\hrulefill
	\vspace*{0pt}
\end{figure*}

\subsection{Spectral Efficiency with the LS Estimator}
If the LS estimator in \eqref{ls1} is used, we obtain a closed-form expression for the SE in \eqref{eq:SEexpression} as in the next theorem.
\setcounter{equation}{22}
   \begin{theorem}
	If MR combining with $\mathbf{v}_{jk} = \frac{1}{\sqrt{p_{jk}} \tau_p} \mathbf{y}^p_{jjk} $ is used based on the LS estimator, then 
	\begin{equation}\label{lseq1}
	\mathbb{E}\left\lbrace \mathbf{v}^H_{jk} \mathbf{h}^j_{jk}  \right\rbrace = {\eta}_{jk}^{\mathrm{ul}}= \mathrm{tr}(\mathbf{R}^j_{jk} )+ \sum_{(l,i) \in \mathcal{P}_{jk} } \frac{\sqrt{p_{li}}}{\sqrt{p_{jk}}} (\bar{\mathbf{h}}^j_{li} )^H \bar{\mathbf{h}}^j_{jk} 
	\end{equation}
	\begin{equation}\label{lseq2}
	\mathbb{E}\left\lbrace \|\mathbf{v}^H_{jk}\|^2 \right\rbrace = \mu^{\mathrm{ul}}_{jk} 	= \frac{1}{p_{jk} \tau_p} \mathrm{tr}\left( (\boldsymbol{\Psi}^j_{jk})^{-1} \right) + \frac{1}{p_{jk} \tau^2_p} \|\bar{\mathbf{y}}^p_{jjk}\|^2 
	\end{equation} 
	and $\mathbb{E}\lbrace |\mathbf{v}^H_{jk} \mathbf{h}^j_{li} |^2 \rbrace$  is defined  in \eqref{lseq3} at the top of this page where $\bar{\mathbf{x}}_{jk} = \bar{\mathbf{y}}^p_{jjk} - \sqrt{p_{li}} \tau_p\bar{\mathbf{h}}^j_{li}$ and $(\boldsymbol{\Omega}^j_{jk})^{-1} = (\boldsymbol{\Psi}^j_{jk})^{-1} -p_{li}\tau_p \mathbf{R}^j_{li}  $. Plugging \eqref{lseq1}, \eqref{lseq2} and \eqref{lseq3} into $\eqref{sec4eq1}$ gives \setcounter{equation}{25}
	\begin{equation}\label{LSeq1} \gamma^{\mathrm{ul}}_{jk} \!=\!
	\frac{p_{jk} \left| {\eta}_{jk}^{\mathrm{ul}}\right| ^2}{ \displaystyle\sum_{l=1}^{L}\sum_{i=1}^{K_l}   p_{li} {\chi}^{\mathrm{ul}}_{li}  -p_{jk}\left| {\eta}_{jk}^{\mathrm{ul}}\right| ^2 + \mu^{\mathrm{ul}}_{jk}  \sigma^2_{\mathrm{ul}} 	 }
	\end{equation}
	where $\chi_{li}^{\mathrm{ul,ls}}$ is defined in \eqref{lseq3} on the top of this page.

\end{theorem}
  \begin{IEEEproof}
	The proof is omitted due to space limitations, but follows from direct, tedious computation of expectations. 
\end{IEEEproof}

Note that the Rayleigh fading counterpart of \eqref{LSeq1} can be easily obtained by setting all the mean values to zero. In this case, the difference in SE between the MMSE and LS estimators is rather small \cite{EmilsBook}. However, the loss in SE incurred by using the LS estimator under Rician fading can be quite large depending on the dominance of the LoS paths. 

Since the mean values are not utilized as prior information, the interference terms are larger than when using the MMSE estimator. The LS estimates of the pilot-contaminating UEs are equal up to a scaling factor. Compared to the SE with MMSE estimator, the inner product of  $\bar{\mathbf{y}}^p_{jjk}$ and $\bar{\mathbf{h}}^j_{li}$ determines how large  the corresponding interference terms are instead of the inner product of $\bar{\mathbf{h}}^j_{jk}$ and $\bar{\mathbf{h}}^j_{li}$.

\section{Numerical Results} \label{sec:numerical_results}
\setcounter{equation}{27}
In this section, the closed-form SE expressions from the previous sections are validated and evaluated by simulating a Massive MIMO cellular network. We have a 16-cell setup where each cell covers a square of $250 \times 250$ m. The network has a wrap-around topology. This layout is employed to guarantee that all BSs receive equally much interference from all directions. There are ten UEs per cell and these are uniformly and independently distributed in each cell, at distances  larger than $35$\,m from the BS. The location of each UE is used when computing large-scale fading and nominal angle between the UE and BSs. Based on the 3GPP model \cite{3gpp}, the large-scale fading for the LoS and NLoS components are model in dB as
\begin{equation}\label{betalos}
	\beta^{j,\mathrm{LoS}}_{li} = -30.18 -26  \log_{10}\left( {d^j_{li}} \right) + F^j_{li}.
\end{equation}
\begin{equation}\label{betanlos}
\beta^{j,\mathrm{NLoS}}_{li} = -34.53 -38  \log_{10}\left( {d^j_{li}} \right) + F^j_{li}.
\end{equation}
where $F^j_{li} \sim \mathcal{N}(  0, \sigma^2_\mathrm{sf} )$ is the shadow fading with standard deviations $\sigma_\mathrm{sf} = 4$ for LoS and $\sigma_\mathrm{sf} = 10$ for NLoS,  and  $d^j_{li}$ is the distance between UE $i$ in cell $l$ and BS $j$ in meters. Each UE is assigned to the BS that provides largest large-scaling coefficient taking shadow fading into account.

 Each BS is equipped with a uniform linear array (ULA) with half-wavelength antenna spacing \cite[Sec.~1]{EmilsBook}, thus the LoS component from UE $i$ in cell $l$ to BS $j$ is
\begin{equation}
\bar{\mathbf{h}}^j_{li} = \sqrt{	\beta^{j,\mathrm{LoS}}_{li}} \left[ 1 \ e^{j\pi \sin(\varphi^j_{li})} \, \dots \, e^{j\pi  (M-1)\sin(\varphi^j_{li})}  \right] ^T
\end{equation}
where ${\varphi}^j_{li} $ is the angle of arrival to the UE seen from the BS.

We use the Gaussian local scattering model to model the covariance matrices \cite[Sec.~2.6]{EmilsBook}, such that
\begin{equation}
\left[ \mathbf{R}^j_{li}\right] _{s,m} = \beta^{j,\mathrm{NLoS}}_{li}  e^{j \pi  (s-m)\sin({\varphi}^j_{li})}  e^{-\frac{\sigma^2_\varphi}{2}\left(\pi (s-m)\cos({\varphi}^j_{li})\right)^2 } 
\end{equation}
where the angular standard deviation $\sigma_\varphi$ determines how the small-scale fading components deviate in angle from the nominal angle. Note that we have purposely not defined any ``$\kappa$-factor'' to keep the model general, but one can compute it for every user using \eqref{betalos} and \eqref{betanlos}.
\begin{figure}[h!]
	\includegraphics[width=0.48\textwidth]{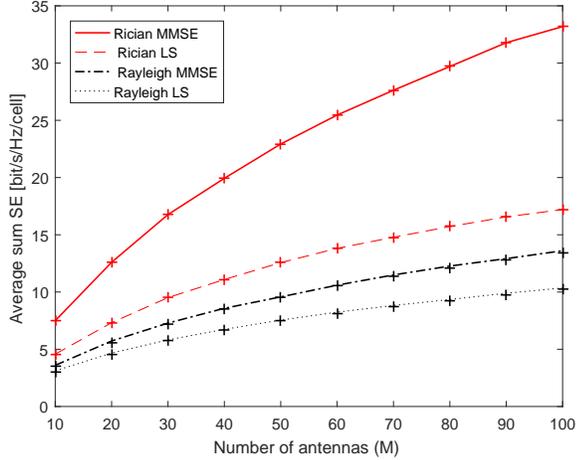}
	\caption{Average UL sum SE for ten UEs as a function of the number of BS antennas for different channel estimators.}  \label{figure1}
\end{figure}

\begin{figure}[h!]
	\includegraphics[width=0.48\textwidth]{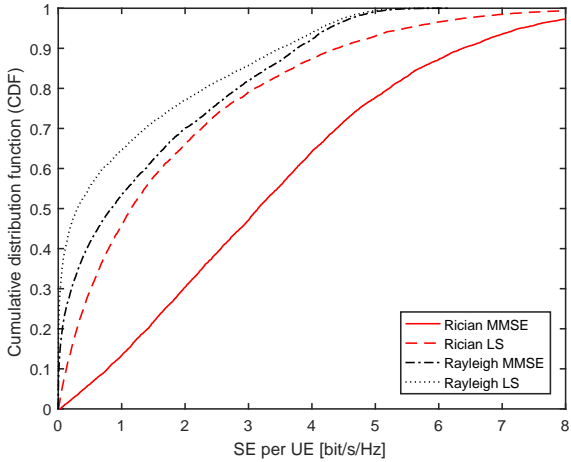}
	\caption{CDF of the UL SE per UE with $M=100$ for different channel estimators. Correlated Rician fading channels are compared with correlated Rayleigh fading channels. } \label{figure2}
\end{figure}
We use $\sigma_\varphi= 10^\circ$ in this simulation. We consider communication over a 20\,MHz channel, the UL transmit power is $10$\,dBm per UE and the total receiver noise power is $-94$\,dBm.  Each coherence block consists of $\tau_c=200$ samples and $\tau_p=10$ pilots are allocated randomly in each cell.

Fig.~\ref{figure1} shows the average sum SE over different UE locations with MR combining when using either the MMSE or LS estimator. As a reference, we also provide curves for Rayleigh fading with the same covariance matrices, representing the case when all the LoS components are blocked but the small-scale fading remains (i.e., the average channel gain $\mathbb{E}\{ \| \mathbf{h}^j_{li}\|^2 \}$ is smaller, as would be the case in practice). The curves are generated using the closed-form expressions and the markers are generated by Monte Carlo simulations. The fact that the markers overlap with the curves confirms the validity of our analytical results. As expected, the SE is higher when the MMSE estimator is employed, since the LoS component and spatial correlation are known. The difference in SE between using the MMSE and LS estimators increases with the number of antennas. 
The difference is smaller with Rayleigh fading since there is no LoS component to learn (the difference will vanish for spatially uncorrelated channels).

Fig.~\ref{figure2} shows cumulative distribution function (CDF) curves for the SE per UE. The randomness is due to random UE locations and shadow fading realizations. 
The main observation is that all of the UEs will (statistically) obtain higher SE when there is a LoS component, since the CDF curves with Rician fading are to the right of the corresponding curves with Rayleigh fading. The choice of estimator has little impact for the UEs with good channels, since the estimation errors are small also with the LS estimator, but MMSE estimation is particularly beneficial for UEs with weak channels. Note that equal power is allocated to all UEs for simplicity. Other power control algorithms, that mitigate near-far effects or optimize the SEs according to specific criteria, are certainly possible but are left for future work.

\section{Conclusion}
 
 This paper studies the UL SE of a multicell Massive MIMO system with spatially correlated Rician fading channels between every BSs and UEs. We derived rigorous closed-form  SE expressions when using either MMSE or LS estimation. The expressions provide exact insights into the operation and interference behavior when having Rician fading channels.
 We observed that the existence of a LoS component improves the achievable UL SE. In addition, the MMSE estimator performs better than the LS estimator for both Rayleigh and Rician fading. However, the covariance matrices and the mean vectors are generally not known perfectly. In practice, the performance lies between the MMSE and LS estimator since it is highly probable that the mean is known up to a random phase-shift and covariance matrices are known with some error.

\bibliographystyle{IEEEtran}
\bibliography{IEEEabrv,references}

\end{document}